# Proximity-induced ferromagnetism in graphene revealed by anomalous Hall effect

Zhiyong Wang, Chi Tang, Raymond Sachs, Yafis Barlas, and Jing Shi

Department of Physics and Astronomy, University of California, Riverside, CA 92521

We demonstrate the anomalous Hall effect (AHE) in single-layer graphene exchange-coupled to an atomically flat yttrium iron garnet (YIG) ferromagnetic thin film. The anomalous Hall conductance has magnitude of ~0.09($2e^2/h$) at low temperatures and is measurable up to ~ 300 K. Our observations indicate not only proximity-induced ferromagnetism in graphene/YIG with large exchange interaction, but also enhanced spin-orbit coupling which is believed to be inherently weak in ideal graphene. The proximity-induced ferromagnetic order in graphene can lead to novel transport phenomena such as the quantized AHE which are potentially useful for spintronics.

Although pristine graphene sheets only exhibit Laudau orbital diamagnetism, local magnetic moments can be introduced in a variety of forms, e.g. along the edges of nanoribbons [1] around vacancies [2] and adatoms [3]. However, a long-range ferromagnetic order in graphene does not occur without exchange coupling between the local moments. In general, introducing local moments and the exchange interaction in bulk materials can be simultaneously accomplished by doping atoms with unfilled *d*- or *f*-shells [4]. For graphene, scattering caused by random impurities could be detrimental to its high carrier mobility, a unique electronic property that should be preserved. By coupling the single atomic sheet of carbons with a magnetic insulator film, e.g. YIG, we may introduce ferromagnetism in graphene without sacrificing its excellent transport properties. The hybridization between the $\pi$-orbitals in graphene and the nearby spin-polarized *d*-orbitals in magnetic insulators gives rise to the exchange interaction required for long-range ferromagnetic ordering. On the other hand, such proximity coupling does not bring unnecessary disorder to graphene. In addition, unlike ferromagnetic metals that could in principle mediate proximity exchange coupling, the insulating material does not shunt current away from graphene. In this work, we demonstrate ferromagnetic graphene via the proximity effect and directly probe the ferromagnetism by measuring the anomalous Hall effect (AHE).

To bring graphene in contact with YIG substrates, we apply a previously developed transfer technique (see SM) that is capable of transferring pre-fabricated functional graphene devices to any target substrates [5]. We first fabricate exfoliated single-layer graphene devices on 290 nm-thick $SiO_2$ atop highly doped Si substrates using standard electron-beam lithography and Au electron-beam evaporation. Both longitudinal and Hall resistivities are measured at room temperature to characterize the state of the pre-transferred devices. To transfer selected devices, we spin-coat the chip with poly-methyl methacrylate (PMMA) followed by a hard bake at 170 °C for 10 minutes. The entire chip is then soaked in 1 M NaOH solution for two days to etch away $SiO_2$ so that the device/PMMA layer is released from the substrate. The PMMA layer attached with the fully nano-fabricated graphene devices is then placed on the target substrate. Finally, the PMMA is dissolved with acetone followed by careful rinsing and drying, and the device is ready for electrical transport and/or Raman measurements. This technique was previously applied to fabricate graphene devices on $SrTiO_3$, a high nominal dielectric constant pervoskite material [5,6]. The transfer steps are schematically shown in Fig. S-1 in SM.

For this study, ~ 20 nm thick atomically flat YIG films are grown epitaxially on 0.5 nm-thick gadolinium gallium garnet (GGG) substrates by pulsed laser deposition as described elsewhere [7], which are then subsequently annealed in an oxygen-flow furnace at 850 °C for 6 hours to minimize oxygen deficiency. Magnetic hysteresis loop measurements and atomic force microscopy (AFM) are performed to characterize the magnetic properties and the morphology of YIG films, respectively. The hysteresis loops of a representative YIG/GGG sample are displayed in Fig. 1(a). The YIG film clearly shows in-plane magnetic anisotropy. The in-plane coercive field and saturation field are both small (~ a few G and < 20 G, respectively), and the out-of-plane loop indicates a typical hard-axis behavior with a saturation field ~2000 G, which can vary from 1500 to 2500 G in different YIG samples. Fig. 1(a) inset shows the AFM topographic image of a typical YIG film. The nearly parallel lines are terraces separated by steps with the atomic height and the roughness on the terrace is ~

0.06 nm. The smoothness of the YIG surface is not only critical to a strong induced proximity effect in graphene, but also favorable for maintaining high carrier mobility [8].

In order to effectively tune the carrier density in graphene/YIG, we fabricate a thin methyl methacrylate (MMA) or PMMA top gate. Fig. 1(b) shows a false-colored optical image of a graphene device on YIG/GGG before the top gate is fabricated. Room-temperature Raman spectroscopy is performed at different stages of the device fabrication. Representative spectra are shown in Fig. 1(c) for the same graphene device on $SiO_2$ (before transfer) and YIG (after transfer), and for YIG/GGG only. Graphene/YIG shows both the characteristic $E_{2g}$ (~1580 cm$^{-1}$) and *2D* peaks (~2700 cm$^{-1}$) of single-layer graphene as well as YIG's own peaks, suggesting successful transfer. We also note that the transfer process does not produce any measurable *D* peak (~1350 cm$^{-1}$) associated with defects [9]. Fig. 1(d) is a schematic drawing of a top-gated transferred device on YIG/GGG.

Low-temperature transport measurements are performed in Quantum Design's Physical Property Measurement System. Fig. 2(a) is a plot of the gate voltage dependence of the four-terminal electrical conductivity scaled by the effective capacitance per unit area, $C_s$. Since different gate dielectrics are used in the back- and top-gated graphene devices, $C_s$ is calculated based on the quantum Hall data which agrees with the calculated value using the nominal dielectric constant and the measured dielectric film thickness. Before transfer, the Dirac point is at ~ -9 V and the field-effect mobility is ~ 6000 cm$^2$/V·s. After transfer, the Dirac point is shifted to ~ -18 V. The slope of the $\sigma_{xx}/C_s$ vs. $V_g$ curve increases somewhat, indicating slightly higher mobility, which suggests that the transfer process, the YIG substrate, and the top-gate dielectric do not cause any adverse effect on graphene mobility. At 2 K, the mobility improves further, exceeding 10000 cm$^2$/V·s on the electron side. Well-defined longitudinal resistance peaks and quantum Hall plateaus are both present at 8 T as shown in Fig. 2(b), another indication of uncompromised device quality after transfer. In approximately 8 devices studied, we find that the mobility of graphene/YIG is either comparable with or better than that of graphene/SiO$_2$.

To study the proximity-induced magnetism in graphene, we perform the Hall effect measurements in the field range where the magnetization of YIG rotates out of plane over a wide range of temperatures. Nearly all graphene/YIG devices exhibit similar nonlinear behavior at low temperatures as shown in Fig. 2(c). Fig. 2(d) only shows the Hall data after the linear ordinary Hall background (the straight line in Fig. 2(c)) is subtracted. In ferromagnets, the Hall resistivity generally consists of two parts [10]: from the ordinary Hall effect and the anomalous Hall effect (AHE), i.e. $R_{xy} = R_H(B) + R_{AHE}(M) = \alpha B + \beta M$, here $B$ being the external magnetic field, $M$ being the magnetization component in the perpendicular direction, and $\alpha$ and $\beta$ are two $B$- and $M$-independent parameters respectively. The $B$-linear term results from the Lorentz force on one type of carriers. Higher order terms can appear if there are two or more types of carriers present. The $M$-linear term is due to the spin-orbit coupling in ferromagnets [10]. The observed non-linearity in $R_{xy}$ suggests the following three possible scenarios: the ordinary Hall effect arising from more than one type of carriers in response to the external magnetic field, the same Lorentz force related ordinary Hall effect but due to the stray magnetic field from the underlying YIG film, and AHE from spin-polarized carriers. The nonlinear Hall curves saturate at $B_s \sim 2300$ G, which is approximately correlated with the saturation of the YIG magnetization in Fig. 1(a). This behavior is characteristic of AHE, i.e. $R_{AHE} \propto M_G$, where $M_G$ is the induced magnetization of graphene. Since $M_G$ results from the proximity coupling with the magnetization of YIG, $M_{YIG}$, both $M_G$ and $M_{YIG}$ should saturate when the external field exceeds some value. The saturation field of YIG is primarily determined by its shape anisotropy, i.e. $4\pi M_{YIG}$, which should not change significantly far below the Curie temperature (550 K) of YIG. On the other hand, if it is caused by the Lorentz force on two types of carriers, the nonlinear feature would not have any correlation with $M_{YIG}$. These experimental facts do not support the first scenario. To further exclude the ordinary Hall effect due to the Lorentz force from stray fields from YIG, we fabricate graphene devices on $Al_2O_3$/YIG, in which the 5 nm thick continuous $Al_2O_3$ layer should have little effect on the strength of the stray field but effectively cut off the proximity coupling. We do not observe any measurable nonlinear Hall signal similar to those in companion graphene/YIG devices (Figs. S-6 and S-7 in SM). It excludes the effect of the stray field. Therefore, we attribute the nonlinear Hall signal in graphene/YIG to AHE which is due to spin-polarized carriers in ferromagnetic graphene. Further evidence will be presented when the gate voltage dependence is discussed below.

Fig. 3(a) shows the AHE resistance, $R_{AHE}$, vs. the positive out-of-plane magnetic field taken from 5 to 250 K. All linear background has been removed. Fig. 3 (b) is the extracted temperature dependence of the saturated AHE resistance. The AHE signal decreases as the temperature is increased, but it stays finite up to nearly 300 K. We note that the AHE magnitude changes sharply in the temperature range of 2 − 80 K, and then stays relatively constant above 80 K before it approaches ~ 300 K, which defines the Curie temperature of $M_G$. In conducting ferromagnets, the AHE resistance, $R_{AHE}$, scales with the longitudinal resistance, $R_{xx}$, in the power-law fashion [10], i.e. $R_{AHE} \propto M_G R_{xx}^n$. Thus the temperature dependence of $R_{AHE}$ could originate from $M_G$ and/or $R_{xx}$. Here $M_G$ should be a slow-varying function of the temperature below 80 K; however, the temperature dependence of $R_{xx}$ in 1T field (inset of Fig. 3(b)) cannot account for the steep temperature dependence of $R_{AHE}$ either. Therefore, we attribute the discrepancy to possible physical distance change between the graphene sheet and YIG either due to an increase in the vibrational amplitude or different thermal expansion coefficients between the top-gate dielectric and YIG/GGG. We have observed variations in both the Curie temperature $T_c$ for $M_G$ and the maximum $R_{AHE}$ (see Fig. S-2 and S-3). Among all 8 devices studied, the highest $T_c$ is ~ 300 K and the largest $R_{AHE}$ at 2 K is ~ 200 Ω.

With a top gate, we can control the position of the Fermi level in graphene at a fixed temperature, not possible in ferromagnetic metals. By sweeping the top-gate voltage, $V_{tg}$, we systematically vary both $R_{AHE}$ and $R_{xx}$ and keep the induced magnetization and exchange coupling strength unchanged. More importantly, by changing the carrier type, a sign reversal occurs in the ordinary Hall, i.e. the slope of the linear background signal. We remove this carrier density dependent linear background for each gate voltage and obtain the AHE signal. Fig. 4(a) is the AHE resistivity of a device measured at 20 K for several $V_{tg}$'s: 60 V (red squares), 0 V (green circles), and -20 V (blue triangles), respectively. The inset shows the $V_{tg}$-dependence of the resistivity. The Dirac point is at ~35 V; therefore, carriers are predominately electrons at 60 V with a density ~ $2.5 \times 10^{11}$ cm$^{-2}$, but predominately holes at both 0 and -20 V. We deliberately avoid the region close to the Dirac point where both electrons and holes coexist and the ordinary Hall signal acquires high-order terms in $B$. In the gate dependence data, it is important to note that the AHE sign remains unchanged regardless of the carrier type. This is strong evidence that the observed nonlinear Hall signal is not due to the ordinary Hall effect from two types of carriers, either from the external or stray field, but due to the AHE contribution from spin-polarized carriers in ferromagnetic sample. In addition, the resistance at 60 V is the highest among the three, followed by that at 0 V, and then -20 V, and the corresponding $R_{AHE}$ magnitude follows the same order.

To further reveal the physical origin of AHE, we now focus on the relationship between $R_{AHE}$ and $R_{xx}$ as $V_{tg}$ is tuned. Fig. 4(b) shows more gate-tuned AHE data in another top-gated device measured at 2 K. We also exclude the data close to the Dirac point (-14 V for this device) for the reason mentioned above. Starting from -10 V, $R_{AHE}$ is the largest. As $V_{tg}$ is increased, the electron density increases, and $R_{xx}$ decreases accordingly, which is accompanied by a steady decrease in $R_{AHE}$. Due to the negatively biased Dirac point, we cannot reach the completely hole-dominated region within the safe $V_{tg}$ range (gate leakage current < 10 nA). On the hole side where the background is still influenced by the two-band transport, we do not observe any evidence of a sign change in $R_{AHE}$. In the inset we plot $R_{AHE}$ vs. $R_{xx}$ as $V_{tg}$ is varied. From the slope of the straight line in the log-log plot, we obtain the exponent of the power-law: $n = 1.9 \pm 0.2$. The same exponent is also obtained in a different gate-tuned device (see Fig. S-4 and S-5). As in many ferromagnetic conductors, the quadratic relationship indicates a scattering-independent AHE mechanism, which is different from the skew scattering induced AHE[10].

It is understood that a necessary ingredient for AHE is the presence of SOC along with broken time reversal symmetry [10]. AHE can result from either intrinsic (band structure effect) or extrinsic (impurity scattering) mechanisms. Haldane showed that for a honeycomb lattice (graphene) the presence of intrinsic SOC (which breaks time reversal symmetry) can lead to quantized AHE (QAHE) for spin-less electrons [11]. Since intrinsic SOC in graphene is very weak (~10 μeV) [12], this effect has not been observed experimentally.

However, an enhanced Rashba SOC is possible when graphene is placed on substrates [13,14] or subjected to hydrogenation [15] due to broken inversion symmetry. Recently, Qiao *et al.* predicted that ferromagnetic graphene with Rashba SOC should exhibit QAHE [16,17]. In this case, the Dirac spectrum opens up a topological gap with magnitude smaller than twice the minimum of exchange and SOC energy scale (see SM). As the Fermi level is turned into the gap, a decrease in the four-terminal resistance is expected along with a simultaneous quantization of the AHE conductivity approaching $2e^2/h$. In devices exhibiting AHE, the largest AHE at 2 K is ~ 200 Ω. Using the corresponding $R_{xx}$ of 5230 Ω, we calculate the AHE contribution and obtain $\sigma_{AHE} \approx 7$ μS $\approx 0.09(2e^2/h)$, nearly one order of magnitude smaller than the predicted QAHE conductivity $2e^2/h$. Clearly we have not reached the QAHE regime due to the intrinsic band structure effect, indicating that the Rashba SOC strength $\lambda_R$ is smaller than the disorder energy scale. From the minimum conductivity plateau, we estimate the energy scale associated with the disorder $\Delta_{dis} = \hbar/\tau \approx 12$ meV, assuming long-ranged Coulomb scattering [18]. Therefore our experimental results suggest that $\lambda_R < 12$ meV. To observe QAHE, it is important to further improve the quality of the devices or to strengthen the Rashba SOC to fulfill $\lambda_R > \Delta_{dis}$, both of which are highly possible.

In order to understand the physical origin of the observed unquantized AHE in our devices, we calculate the intrinsic AHE (see SM) at the relevant densities for $\lambda_R < 12$ meV. Our results show that the intrinsic AHE conductivity at these densities is an order of magnitude smaller than the observed value, which argues against the intrinsic mechanism. Since charged impurity screening in graphene becomes extremely weak as the Dirac point is approached, it is likely that the extrinsic mechanisms play a more important role here. We would like to point out that gate tunability in ferromagnetic graphene allows for the observation of Fermi energy dependence of the AHE conductivity, which cannot be achieved in ordinary ferromagnet metals. If the carrier density can be modulated by gating, besides the exponent, the Fermi energy dependence of the AHE conductivity can be experimentally determined over a broad range of energy [19]. This additional information can help further pinpoint the physical origin of AHE in 2D Dirac fermion systems.

We thank Z.S. Lin, T. Lin, B. Barrios, Q. Niu, and W. Beyerman for their help and useful discussions. ZYW and JS were supported by the DOE BES award #DE-FG02-07ER46351, CT was supported by NSF/ECCS, and RS was supported by NSF/NEB.

FIG. 1. (a) Magnetic hysteresis loops in perpendicular and in-plane magnetic fields. Inset is the AFM topographic image of YIG thin film surface. (b) Optical image (without top gate) and (d) schematic drawing (with top gate) of the devices after transferred to YIG/GGG substrate (false color). (c) Room temperature Raman spectra of graphene/YIG (purple), graphene/$SiO_2$ (red), and YIG/GGG substrate only (blue).

FIG. 2. (a) The gate voltage dependence of the device conductivity scaled by the capacitance per unit area for the pre-transfer (293 K, black) and transferred devices (300 K, red; 2 K, green) with the same graphene sheet. (b) Quantum Hall effect of transferred graphene/YIG device in an 8 T perpendicular magnetic field at 2 K. (c) The measured total Hall resistivity data at 2 K with a straight line indicating the ordinary Hall background. (d) The nonlinear Hall resistivity after the linear background is removed from the data in (c).

FIG. 3. (a) AHE resistance at different temperatures. (b) The temperature dependence of AHE resistance. Inset is the longitudinal resistance at the Dirac point with no magnetic field (black) and a 1 T perpendicular magnetic field (red).

FIG. 4. (a) AHE resistance with different carrier types and concentrations at 20 K. Inset, gate voltage dependence at 20 K. Red squares, green circles, and blue triangles represent 60 V, 0 V, -20 V top gate voltages, respectively. The sharp noise-like field-dependent features are reproducible. (b) Top gate voltage dependence of the AHE resistance at 2 K. Inset is the log-log plot of $R_{AHE}$ vs. $R_{xx}$. Red curve is a linear fit with a slope of $1.9 \pm 0.2$.

FIG. 1

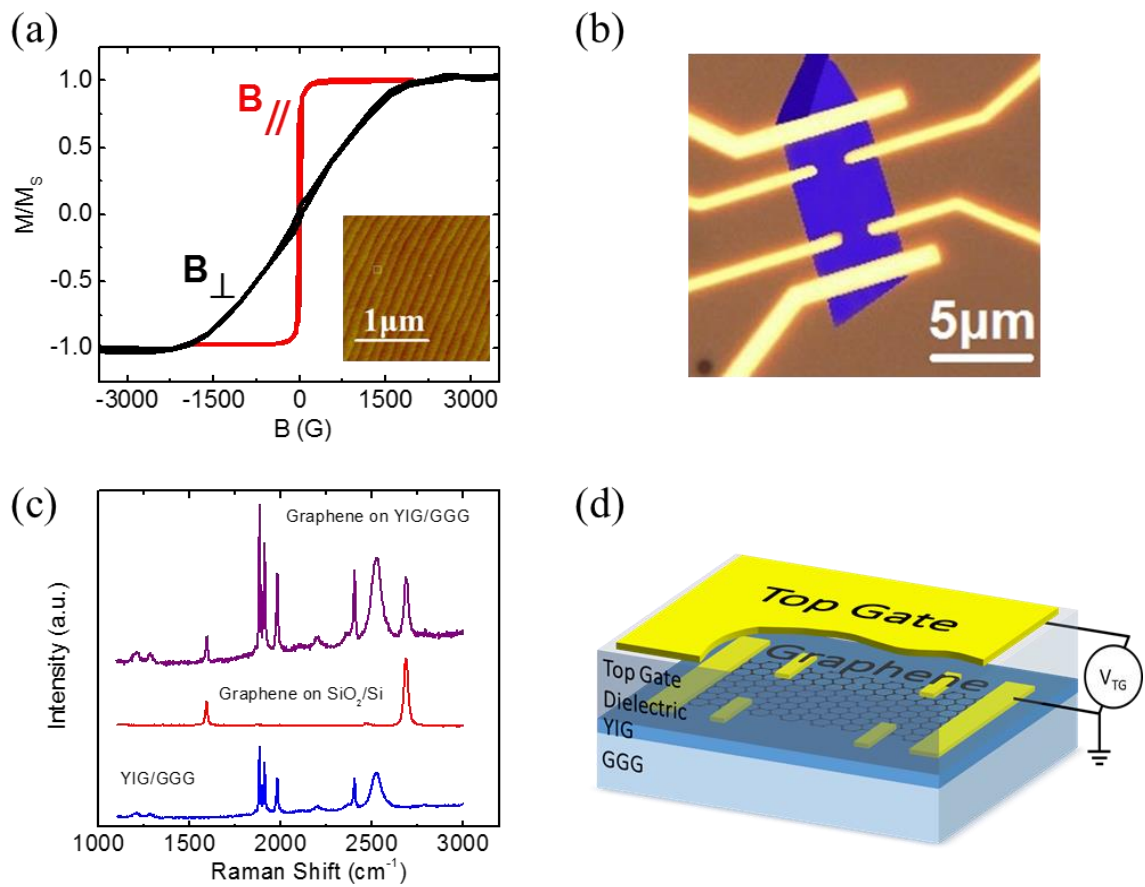

FIG. 2

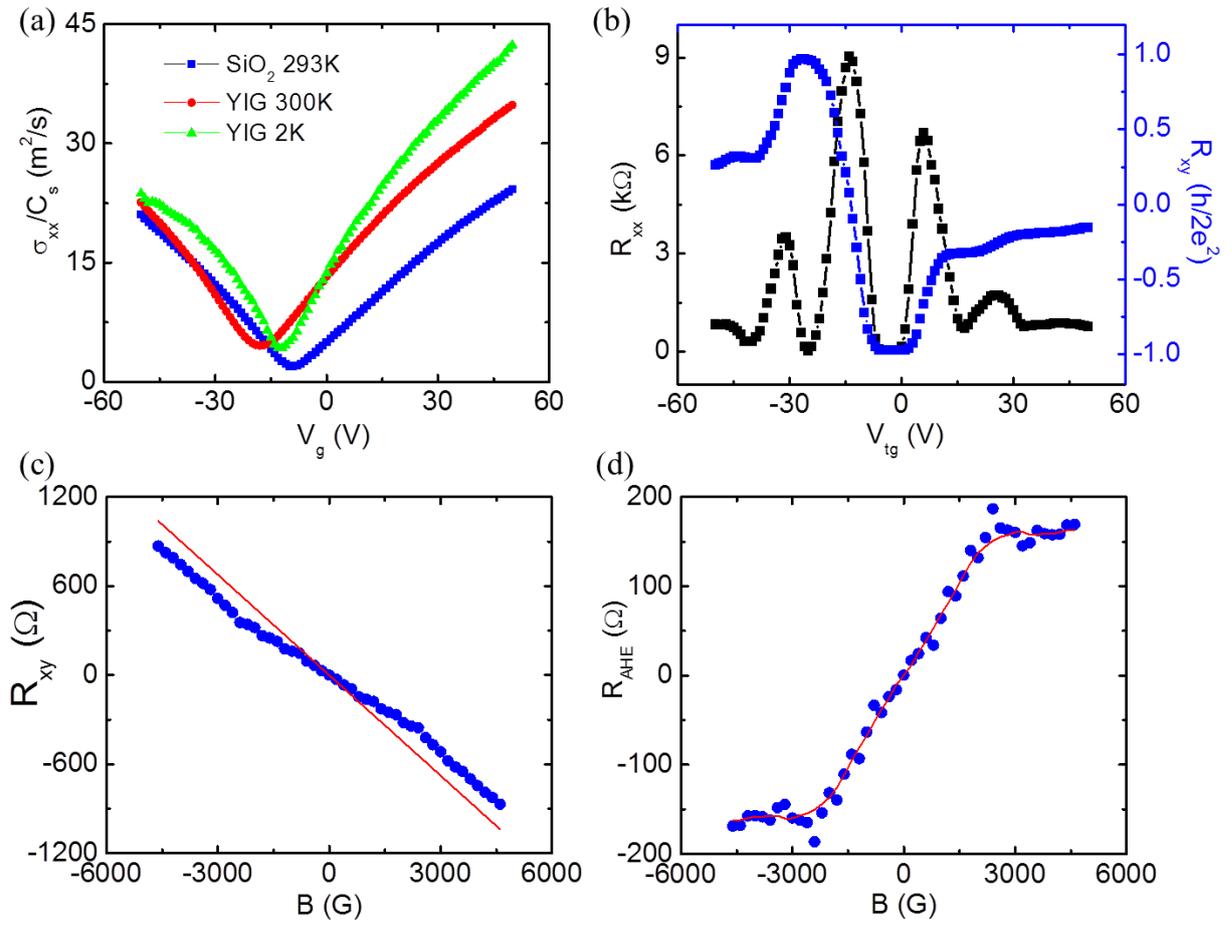

FIG. 3

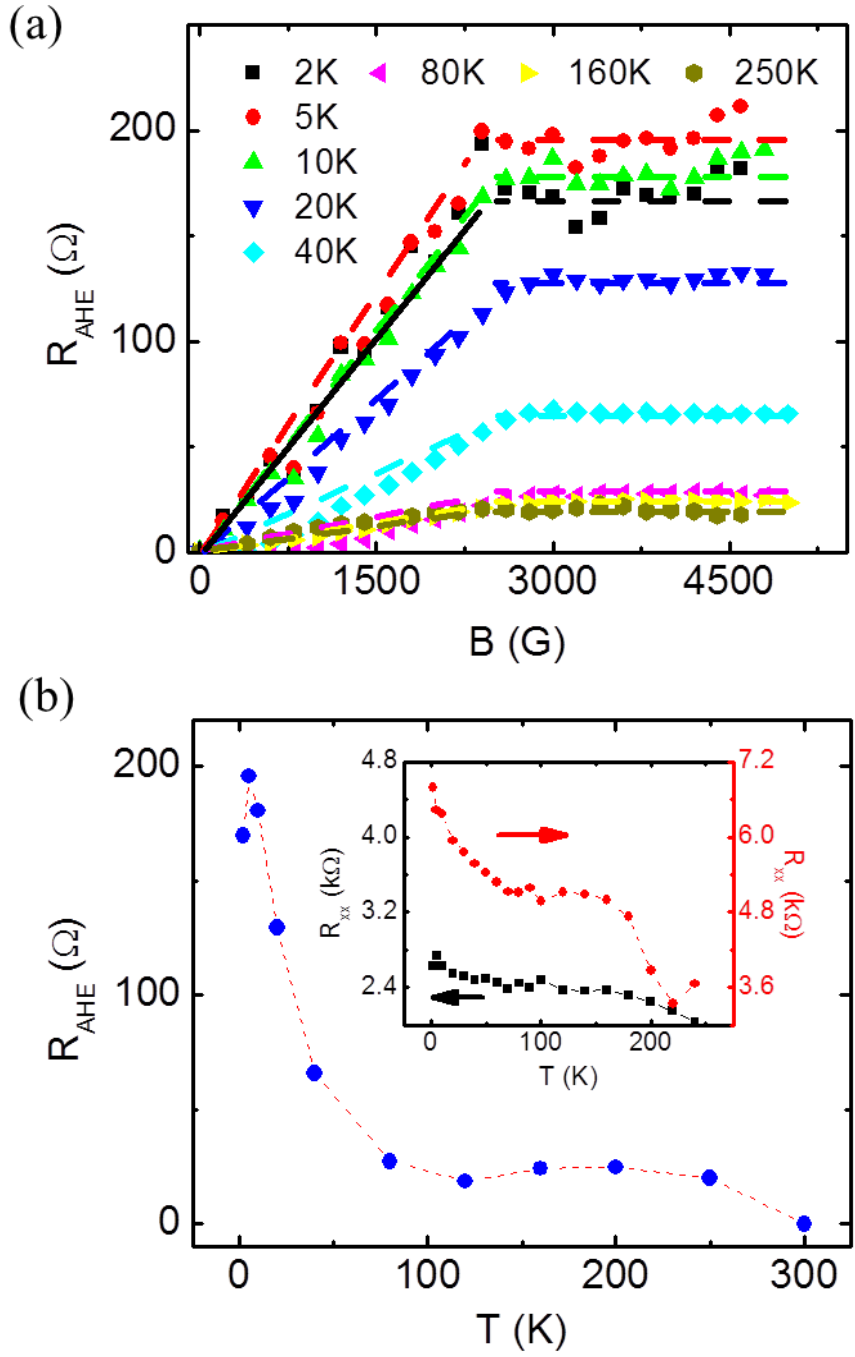

FIG. 4

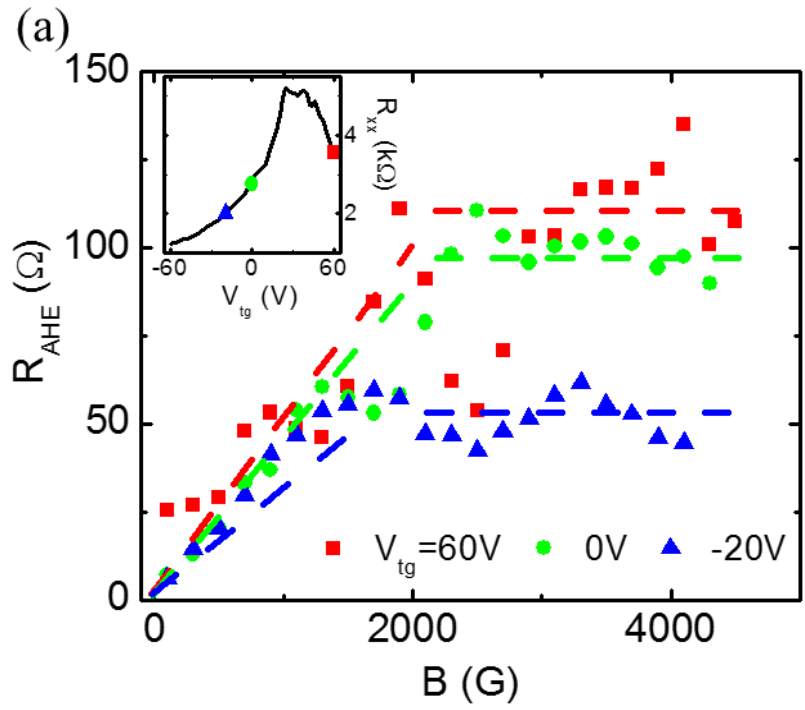

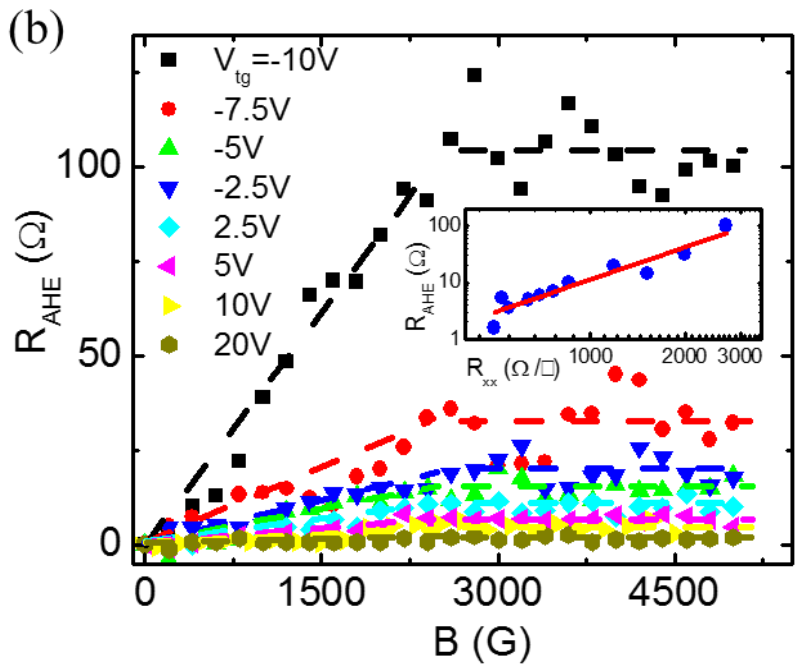